\documentclass[twocolumn,showpacs,preprintnumbers,amsmath,amssymb,pre]{revtex4}
\usepackage{color}

\usepackage{graphicx}
\usepackage{dcolumn}
\usepackage{bm}
\usepackage{subfigure}
\usepackage{amssymb}
\usepackage{multirow}
\graphicspath{{plots/}}

\begin{document}

\title{Diffusion in a strongly coupled magnetized plasma}

\author{T. Ott}
\affiliation{%
    Christian-Albrechts-Universit\"at zu Kiel, Institut f\"ur Theoretische Physik und Astrophysik, Leibnizstra\ss{}e 15, 24098 Kiel, Germany
}%
\author{M. Bonitz}%
\affiliation{%
    Christian-Albrechts-Universit\"at zu Kiel, Institut f\"ur Theoretische Physik und Astrophysik, Leibnizstra\ss{}e 15, 24098 Kiel, Germany
}%

\date{\today}

\begin{abstract}
A first-principle study of diffusion in a strongly coupled one-component plasma in a magnetic field $B$ is presented. As in a weakly coupled plasma, the diffusion coefficient perpendicular to the field exhibits a Bohm-like $1/B$-behavior in the strong-field limit, but its overall scaling is substantially different. The diffusion coefficient parallel to the field 
is strongly affected by the field as well and also approaches a $1/B$ scaling, in striking contrast to earlier predictions.
% of a constant value for large $B$.
\end{abstract}

\maketitle

The question of how a magnetic field influences the transport properties of ensembles of charged particles is of considerable 
significance for a wide spectrum of physical systems and situations. Of particular current interest are the diffusion properties in the case of strong coupling, i.e., when the Coulomb interaction exceeds the kinetic energy of the particles. In astrophysics, for example, the diffusion of strongly coupled ions directly influences the age estimates of white dwarf stars through the timescale of gravitational energy release~\cite{Hughto2010}. Also, knowledge of the diffusion coefficient of strongly correlated nuclei is a key to understand the properties of the crust of magnetized neutron stars~\cite{peng07}. Furthermore, the diffusion coefficient is directly connected to the energy loss (stopping power) of slow ions in dense plasmas~\cite{Dufty1995,Deutsch2009} and, therefore, of special interest for magnetized target fusion scenarios~\cite{Basko2000} as well as in ion beam cooling setups~\cite{Nersisyan2003}. Finally, recent experimental advances with ions in traps as well as with
dusty plasmas~\cite{Lipaev2007,Land2010,Bonitz2010a} are starting to access the microscopic motion of correlated charge carriers in a strong magnetic field. 
A second line of research originates in the comparatively younger field of strongly coupled plasmas (SCP). Here, correlations between particle movements give rise to many new effects including non-Fickian diffusion in two-dimensional geometries~\cite{Ott2008,Ott2009b} or shear waves~\cite{Golden2000}. While the mass transport in an unmagnetized SCP has been investigated in detail,~e.g. Refs.~\cite{Hansen1975,Daligault2006}, only little is known about transport in plasmas which are {\em both strongly coupled and magnetized} which is the subject of the present paper.

Recall that diffusion can be understood as a stochastic process where particles advance in space on average a distance $\Delta r$ until they undergo a scattering event 
with mean frequency $\nu$, giving rise to the diffusion coefficient
\begin{equation}
 D = (\Delta r)^2 \nu.
\label{eq:d_dif}
\end{equation}
A strong magnetic field $B$ substantially alters diffusion of a weakly correlated one-component plasma (OCP). The cross-field diffusion coefficient is obtained by replacing $\Delta r$ by the  Larmor radius $r_c\sim B^{-1}$ and using for $\nu$ the collision frequency. This gives rise to the well-known scaling  of $D_{\perp}\sim B^{-2}$ \cite{lali10} which also applies to non-neutral confined plasmas if multiple collisions are included~\cite{Dubin1997}. A different scaling, $D_{\perp}\sim B^{-1}$, was predicted first by Bohm (``Bohm diffusion'') and plays a key role in magnetic fusion plasmas, e.g. Ref.~\cite{spitzer}. More generally, this ``anomalous'' diffusion is expected to be dominant in plasmas with strong field fluctuations, instabilities or plasma turbulence, as was shown based on mode coupling theory by Marchetti et al.~\cite{Marchetti1984}. An overview of the predicted scalings is given in Tab.~\ref{tab:summary}.
In contrast to weakly coupled plasmas where {\em diffusion along ${\bf B}$} is unaffected by the magnetic field, 
in the case of strong correlations particle motion perpendicular and parallel to {\bf B} become coupled.
Suttorp and Cohen have carried out an extensive analysis based on the Balescu-Guernsey-Lenard (BGL) kinetic equation and predicted 
that, for strong fields, $D_{\parallel}$ would saturate at two-thirds of the field-free value $D_0$, for arbitrary coupling~\cite{Cohen1984}. 
\begin{table*}[ht]
 \centering
\renewcommand{\arraystretch}{1.4}
\begin{tabular}{lcc|cc}
\toprule
&\multicolumn{2}{c}{weak coupling } &\multicolumn{2}{c}{strong coupling }\\[1ex]

& $\beta \lesssim 1$ & $\beta \gg 1$ & $\beta \lesssim 1$  & $\beta \gg 1$ \\
\hline

$D_\perp \phantom{xxxx}$ & $b_0 B^0 +b_2 B^{-2}$ \cite{lali10} \phantom{xx}& $ \gamma_0 k_BT(qB)^{-1}$ \cite{spitzer,Marchetti1984} & & $\to 0$ \cite{Ranganathan2003}\\

&   &  $\sim B^{-2}$ \cite{lali10,Dubin1997} &  ${\bf B^{-\alpha_{\perp}}}$ &  ${\bf \gamma_{\perp} k_BT(qB)^{-1}}$  \\ 

\hline
$D_\parallel $ &  $ \sim B^{0}$ \cite{lali10} &   $\sim B^{0}$ \cite{Cohen1984} & \phantom{xxxx}$( a_0 B^{0} + a_2 B^{2})^{-1}$ \cite{Cohen1984} \phantom{xx}& $\sim B^{0}$ \cite{Cohen1984}, $\to 0$ \cite{Ranganathan2003}\\
& & & ${\bf B^{-\alpha_{\parallel}}}$ &  ${\bf \gamma_{\parallel} k_BT(qB)^{-1}}$ \\
\hline
\end{tabular}
\caption{\label{tab:summary} Reported B-field scalings of the cross-field and parallel diffusion coefficient of an OCP. Results of this work for 
strong coupling, $\Gamma \ge 1$, are in bold, see also Tab.~\ref{tab:decay}. $\beta=\omega_c/\omega_p$, and $a_k,b_k, \alpha_k, \gamma_k$ are $B$-independent coefficients.} 
\end{table*}
Finally, there exist molecular dynamics (MD) simulation results for a SCP by Bernu which, however, are limited to three values of $B$
and do not allow for a conclusive deduction of scaling laws~\cite{Bernu1981}. 
A more recent MD study~\cite{Ranganathan2003} reported an unexpected non-uniform decay of the diffusion coefficients with B and a rapid breakdown 
of diffusion for large $B$~\cite{Note2}. 

In this work, we report on extensive novel first-principle simulations 
of diffusion of magnetized strongly coupled three-dimensional OCPs in a broad range of parameters with the following main results (see also Tab.~\ref{tab:summary}):
1) At large $B$, a Bohm-like scaling of $D_{\perp}$ exists at all couplings,
2) $D_{\parallel}$ exhibits a Bohm-like scaling as well but only in the strongly correlated fluid regime,
3) for moderate coupling the scaling of $D_\parallel$ is slower than $1/B$, however, in all cases the decay 
continues algebraically at strong fields and does not saturate as predicted in Ref.~\cite{Cohen1984}. 
These results are of direct relevance for all SCP that are well described by the OCP model.

\begin{figure}[t]
 \includegraphics[scale=0.45]{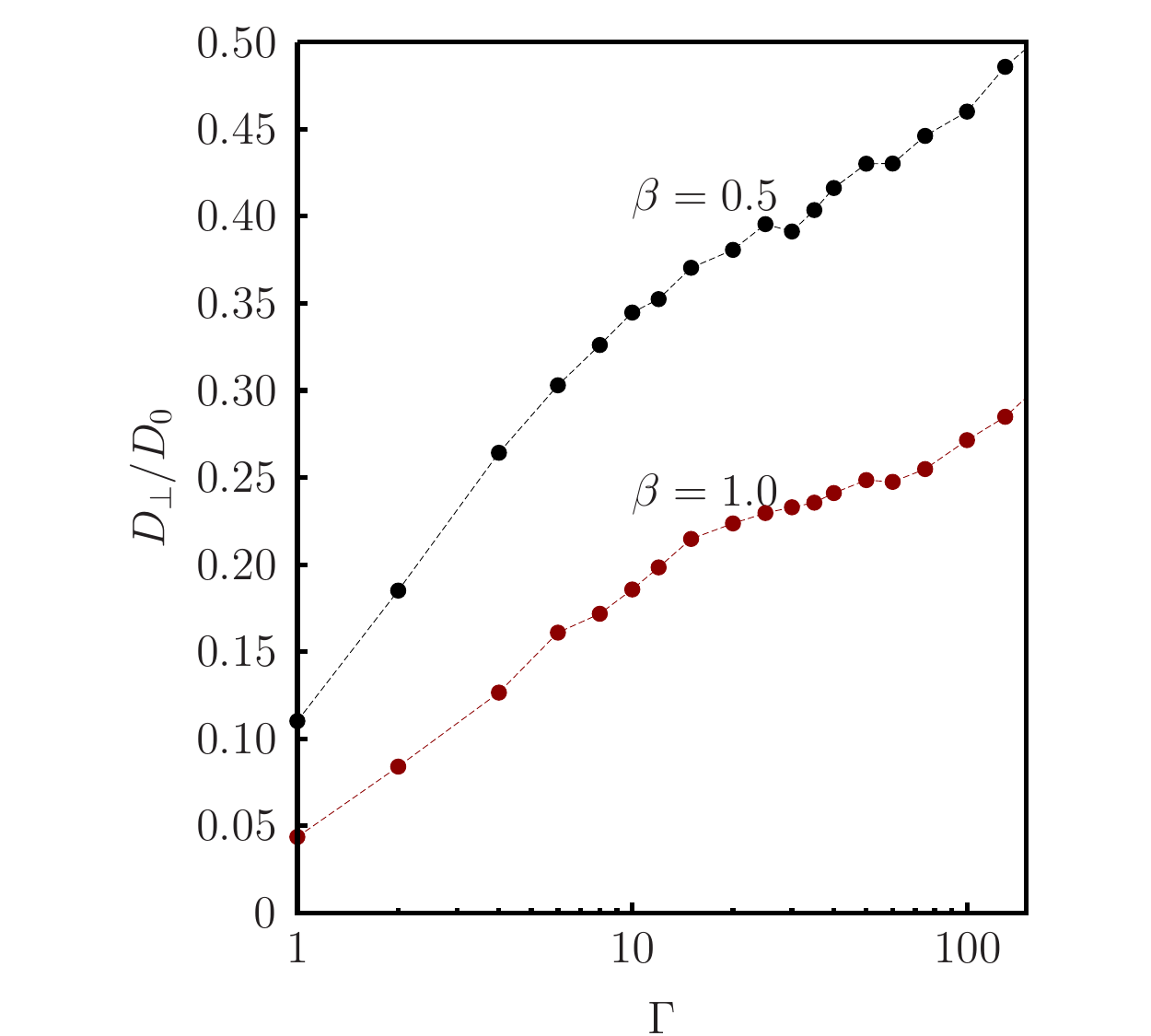}
\caption{(color online) $D_\perp$ in units of its field-free value (for the same $\Gamma$) for two field strengths, $\beta=0.5$ and $1.0$. } 
\label{fig:ratio_perp}
\end{figure}

\begin{figure}[t]
 \includegraphics[scale=0.55]{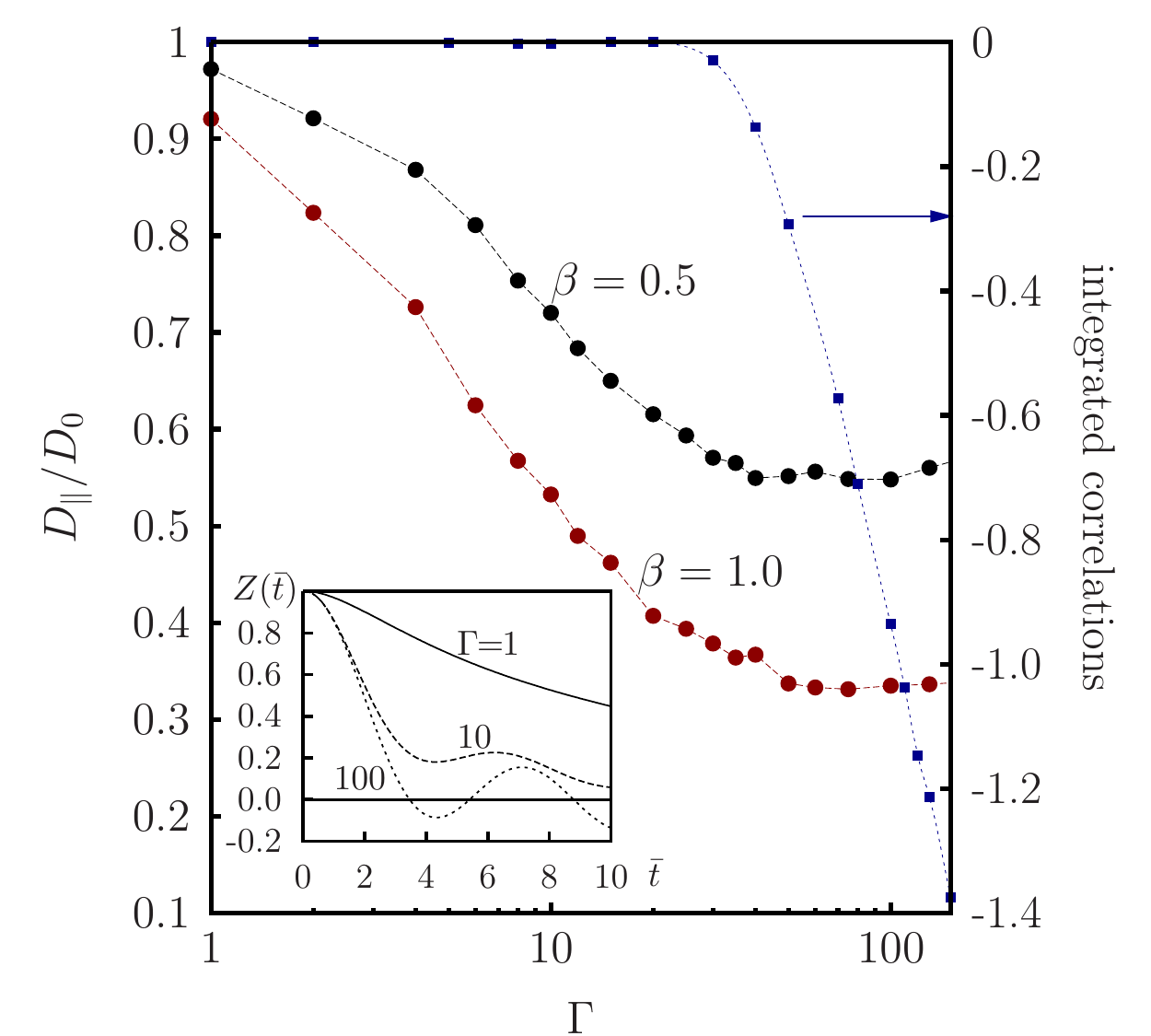}
\caption{(color online)  $D_\parallel$ in units of its field-free value (for the same $\Gamma$) for $\beta=0.5$ and $1.0$.
Also shown are the integrated negative velocity auto-correlations (caging time, see text). Inset: The velocity autocorrelation function $Z(\bar t)$
for three values of $\Gamma$. 
} 
\label{fig:ratio_para}
\end{figure}

We now turn to the description of our results. The thermodynamic state of the OCP is characterized by the coupling parameter~$\Gamma=q^2\times (4\pi\varepsilon_0 ak_BT)^{-1}$, 
where $a=\left[3/\left (4 n \pi \right)  \right]^{1/3}$ is the Wigner-Seitz-radius with the number density $n$, and $q$ and $m$ denote particle charge and mass. The strength of
the magnetic field $B$  (${\bf B} \parallel  {\bf e}_z$) is given by $\beta=\omega_c/\omega_p \propto B$, the ratio of the cyclotron frequency, $\omega_c = qB/m$, 
and the plasma frequency, $\omega_p^2= n q^2 /\left( \varepsilon_0 m\right)$. In the following, we use reduced units $\bar r=r/a$ and
$\bar t=t\omega_p$ for lengths and time. 
The equations of motion, 
\begin{equation}
\ddot{\vec r}_i = \vec F_i/m + \omega_c\,\dot{\vec r}_i \times \hat{\vec {e_z}}, \quad i=1\dots N,
\label{eq:eom}
\end{equation}
where $F_i$ is the Coulomb force due to all particles $j\ne i$,
%
%Eq.~\eqref{eq:eom} 
are solved for $N=8196$ particles by standard MD techniques with Ewald summation and integrators adopted to the influence of the magnetic field~\cite{Spreiter1999,Chin2008}. Periodic boundary conditions are imposed in all three directions of the cubic simulation box. Prior to measurement, the system is brought into equilibrium by a repeated rescaling of the particles' momenta towards the target value of $\Gamma$. After that the system is advanced according to Eq.~\eqref{eq:eom}, resulting in a microcanonical propagation. 
%
%\begin{figure*}
\begin{figure*}[ht]
 \includegraphics[scale=0.6]{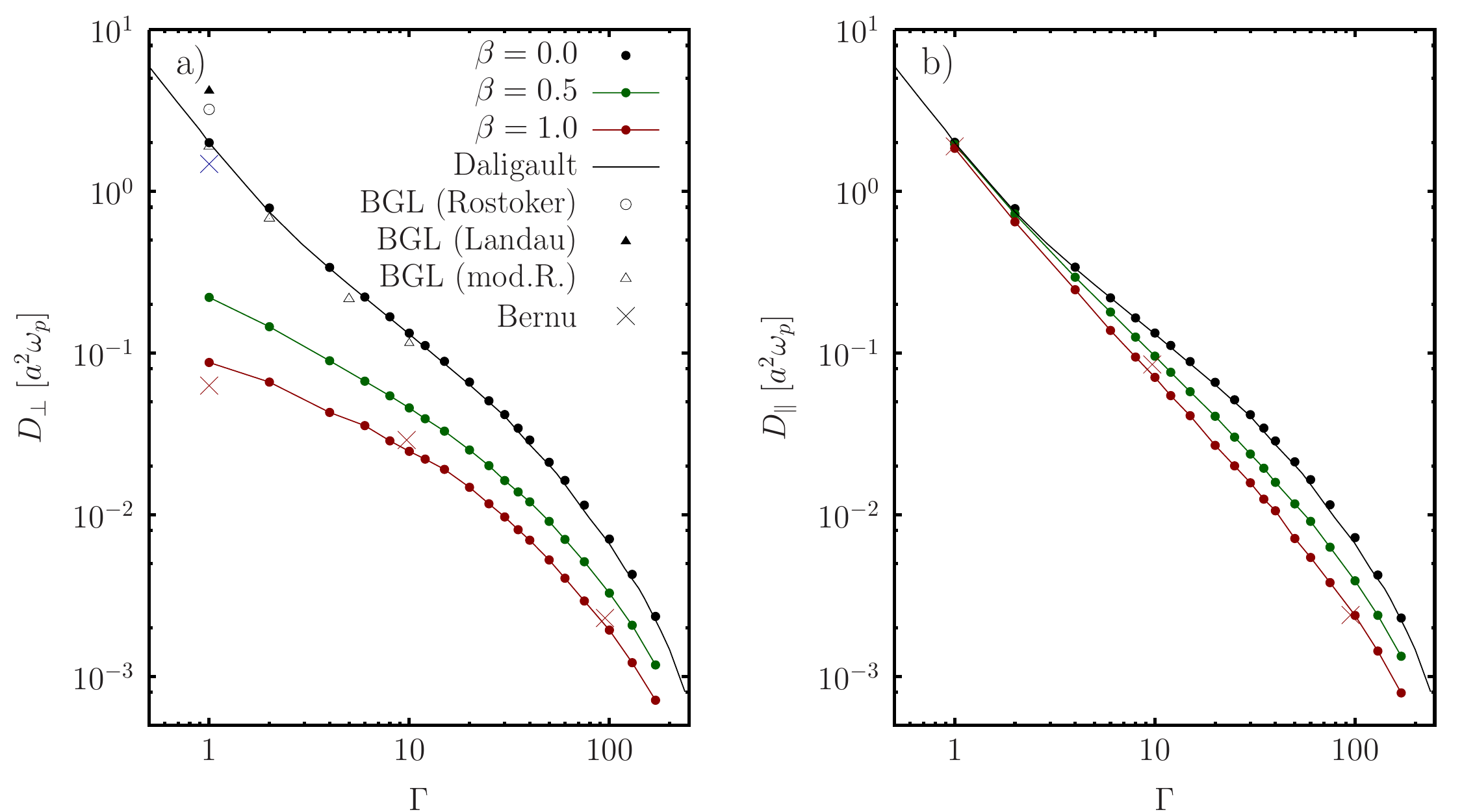}
\caption{(color online) Diffusion coefficient a) perpendicular and b) parallel to the magnetic field 
as a function of coupling for the unmagnetized system and two moderate values of $B$. 
Also shown are the results of Daligault~\cite{Daligault2006} (solid black line), Bernu~\cite{Bernu1981} (crosses) and from kinetic theory (BGL \cite{Cohen1984}).
} 
\label{fig:dvsg}
\end{figure*}
\begin{figure*}
 \includegraphics[scale=0.6]{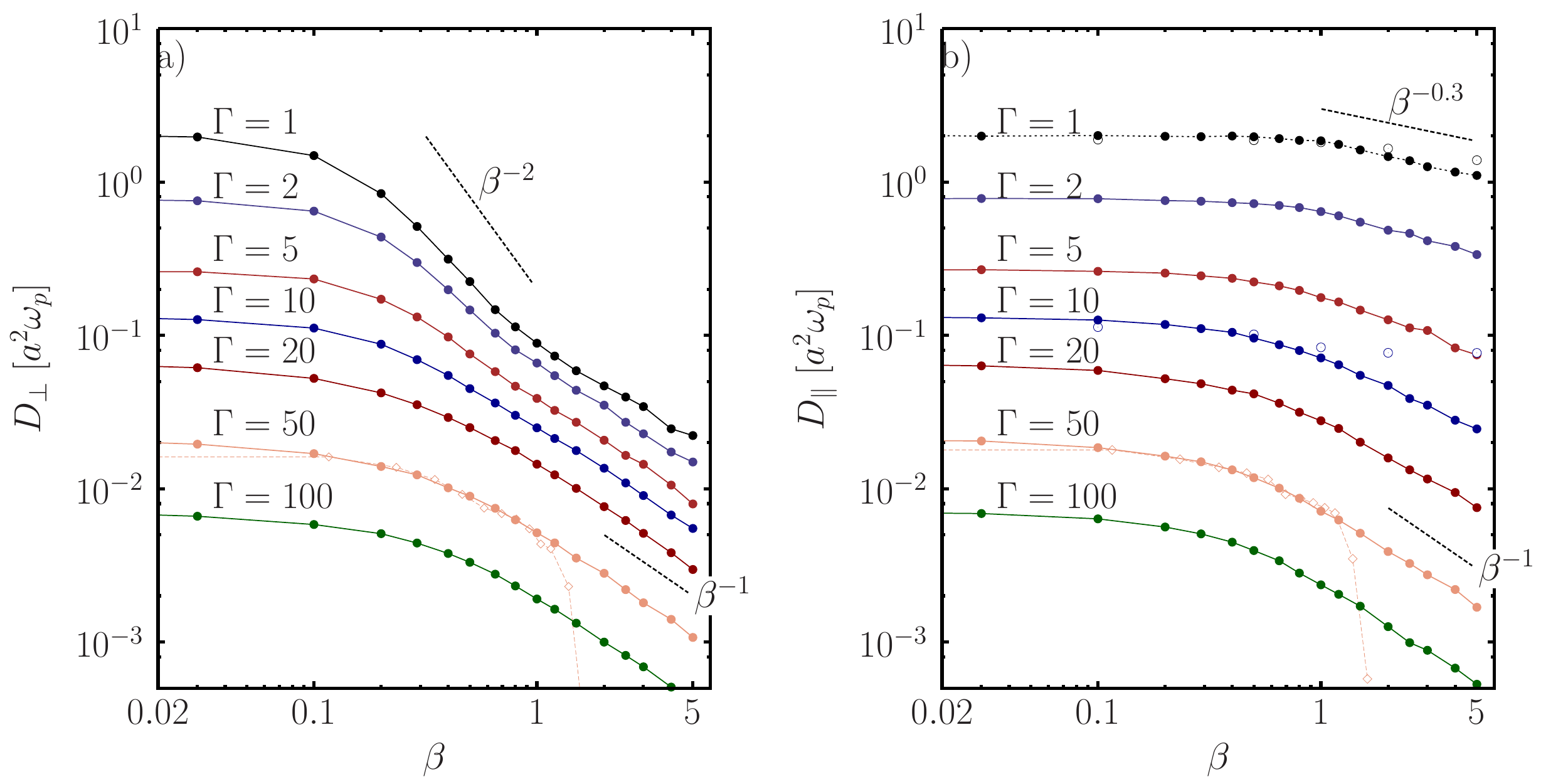}
\caption{(color online)  Diffusion coefficient a) perpendicular and b) parallel to the magnetic field versus field strength $\beta=\omega_c/\omega_p$ for different $\Gamma$. 
The straight broken lines indicate a decay of $\beta^{-2}$, $\beta^{-1}$, and $\beta^{-0.3}$, respectively. Also shown are simulation data from Ref.~\cite{Ranganathan2003} for $\Gamma=50$ (open diamonds) \cite{Note2}. The open circles in b) are predictions from BGL theory with a modified Rostoker Kernel for $\Gamma=1$ and $10$~\cite{Cohen1984}. 
} 
\label{fig:dvsb}
\end{figure*}
From the equilibrium dynamics the diffusion coefficients are calculated by the Einstein relation for the mean-squared displacements~\cite{Hansen2006}
\begin{eqnarray}\label{eq:einstein_perp}
 D_\perp &=& \lim_{t\rightarrow \infty} \frac{\left \langle \left\vert x(t)-x(t_0)\right\vert^2 \right\rangle+\left \langle \left\vert y(t)-y(t_0)\right\vert^2 \right\rangle}{4 t}\,,\\
 D_\parallel &=& \lim_{t\rightarrow \infty} \frac{\left \langle \left\vert z(t)-z(t_0)\right\vert^2 \right\rangle}{2 t}
\label{eq:einstein_para},
\end{eqnarray}
where $\langle \dots \rangle$ denotes averaging over the entire particle ensemble~\cite{Note1}. We verified that the results obtained from Eqs.~(\ref{fig:ratio_perp},\ref{fig:ratio_para}) agree, within the statistical uncertainty, with the time integral over the   velocity auto-correlation function ${Z(t)=\langle \vec v(t)\cdot \vec v(t_0)\rangle}/{N_d}$, (Green-Kubo relation~\cite{Hansen2006}).

Consider first the results for the cross-field diffusion coefficient $D_\perp$. Increase of   
$\Gamma$ leads to a rapid monotonic reduction of $D_\perp$, cf. Fig.~\ref{fig:dvsg}a. Similarly,  a monotonic reduction of $D_\perp$ with $B$ is observed, as expected from weak-coupling arguments. The agreement with the results of Daligault~\cite{Daligault2006} ($\beta=0$) and Bernu~\cite{Bernu1981} is excellent for $B=0$ and fair for magnetized systems. We also include kinetic theory
results from Ref.~\cite{Cohen1984} for different approximations for the memory kernel, nameley Landau, Rostoker and modified Rostoker. While deviations for the former two are considerable, the latter turns out to be fairly accurate, for $B=0$. 

In Fig.~\ref{fig:dvsb}a we detail the depencde $D_\perp(\beta)$ at different $\Gamma$. For $\Gamma=1$, we observe $B$-independence, for small $\beta$, with a crossover to a $\beta^{-2}$-decay around $\beta=0.1$. For large $\beta$ the decay of  $D_\perp$ approaches a $\beta^{-1}$ asymptotics, as observed for weak coupling in Ref.~\cite{Marchetti1984}, see also \cite{Deutsch2009}. Consider now the results for strong coupling, $\Gamma=2\dots 100$.
Irrespective of the coupling, $D_\perp$ is independent of $\beta$ for small values of $\beta$. 
With increasing $\Gamma$ the intermediate $\beta^{-2}$ decay is gradually lost. 
Finally, at high couplings ($\Gamma \gtrsim 10$), a Bohmian regime $\sim \beta^{-1}$ immediately follows for $\beta \gtrsim 0.5$. 
\begin{table}
 \centering
\renewcommand{\arraystretch}{1.1}
\begin{tabular}{cccc}
\hline\hline
$\Gamma$ & \multicolumn{2}{c}{$D_\perp $}  & $D_\parallel$\\
$ \phantom{x} $&\phantom{x} $[\beta=0.2-0.5]$\phantom{x} & \phantom{x}$[\beta=1.0-5.0]$\phantom{x} & \phantom{x}$[\beta=1.0-5.0]$ \\\hline
$1$ &$1.41$& $0.89$ & $0.34$  \\ 
$2$ &$1.15$& $0.92$ & $0.40$  \\ 
$5$ &$0.86$& $0.93$ & $0.52$  \\ 
$10$ &$0.71$ & $0.94$ & $0.6$3  \\ 
$20$ &$0.56$& $0.94$ & $0.80$  \\
$50$ &$0.46$& $0.94$ & $0.86$  \\  
$100$ &$0.45$ & \phantom{xxiiiiii}$0.95\; (\textit{0.99})$ & \phantom{xxiiiiii}$0.94\; (\textit{0.95})$  \\ 
\hline
\end{tabular}
\caption{\label{tab:decay} The modulus $\vert\alpha\vert$ of the decay exponent of $D_\perp$ and $D_\parallel$ ($\sim \beta^{-\alpha}$)
averaged for different ranges of $\beta$. The values in parantheses correspond to the region $[\beta=5.0-100.0]$. } 
\end{table}
Our findings for the decay exponent $\alpha$ ($D_\perp\sim\beta^{-\alpha}$) are collected in Table~\ref{tab:decay} for different ranges of the magnetic field. 
%At intermediate values $\beta=0.2-0.5$ and $\Gamma=1;2$, the decay is considerably faster than $\alpha=1$, while at high coupling, $\Gamma\geq 5$, we find $0<\alpha<1$ (the ``cross-over'' between the two regimes). At strong magnetic field, $\beta>1$, $\alpha$ is close to unity indicating approach to Bohmian diffusion. 
%
Finally, we depict in Fig.~\ref{fig:ratio_perp} the ratio of $D_\perp$ to 
the field-free coefficient $D_0$ for $\beta=0.5$ and $1.0$. At small $\Gamma$, cross-field diffusion is efficiently suppressed by the magnetic field. As, however, $\Gamma$ increases, the increased collisionality of the plasma allows for a more frequent crossing of the magnetic field lines and the diffusion coefficient attains 
a sizeable fraction of its field-free value. 
%The exact value of this fraction depends, of course, on the absolute strength of the magnetic field. 

The present results are, to our knowledge, the first observation of Bohmian diffusion in a SCP. Such a behavior is remarkable because it indicates striking similarities of these highly ordered systems with the completely different turbulent fusion plasmas, despite the entirely different range of coupling parameters: in both cases 
the dominant transport mechanism are collective modes. In an SCP these are predominantly magnetoplasmons with a frequency around ${\tilde \omega}_p \sim (\omega_p^2 + \omega_c^2)^{1/2}$. Indeed, replacing, in Eq.~(\ref{eq:d_dif}), $\Delta r \to r_c$ and $\nu \to {\tilde \omega}_p$ yields $D_{\perp} \to \gamma_{\perp}(\Gamma) k_BT (qB)^{-1}$ for $\beta \gg 1$. For example, for $\Gamma=100$ we find $\gamma_\perp\approx 2/3$ which is surprisingly close to Spitzer's weak coupling estimate of $\gamma_0=1/2.2$ ~\cite{spitzer} but substantially larger than the familiar $1/16$ of Bohm.

Let us now turn to the {\em field-parallel diffusion}. As expected, for small couplings, $D_\parallel$ is practically insensitive to $B$,  (Fig.~\ref{fig:dvsg}b). In contrast, in an SCP in the liquid state that sustains shear, $D_\parallel$ decreases with increasing $\beta$. 
This effect is elucidated in more detail in Fig.~\ref{fig:ratio_para} showing the ratio $D_\parallel/D_0$ as a function 
of $\Gamma$ for two values of $\beta$. At small $\Gamma$, the ratio is close to unity and declines 
steadily with growing $\Gamma$ until, around $\Gamma_\textrm{crit}\approx 30$, it becomes $\Gamma$-independent [in Fig.~\ref{fig:dvsb}.b this corresponds to parallel curves for different $\Gamma$]. Interestingly, the onset of this saturation coincides with a change in the particle dynamics: at
$\Gamma_\textrm{crit}$, the velocity autocorrelation function $Z(t)$ [see inset of Fig.~\ref{fig:ratio_para}] changes from a modulated decay to large amplitude 
oscillations which extend to negative values [we demonstrate this by showing, in Fig.~\ref{fig:ratio_para}, the integral over the negative portions of $Z(t)$ from $\bar t=0$ to $\bar t=100$]. Such negative correlations reflect ``caging'' of particles 
in their local potential minima which lasts on average a time $T_c$. Donk{\'o} \emph{et al.} showed that, in this regime, $D\sim T_c^{-1}$~\cite{Donk'o2002}. 
Thus, Fig.~\ref{fig:ratio_para} indicates that in an external $B$ field parallel diffusion for $\Gamma \gtrsim \Gamma_{\rm crit}$ is limited by caging effects whre 
the magnetic field prolongs $T_c$. 
%in a multiplicative manner, $T_c \rightarrow f(\beta)T_c$. 

The $\beta$-dependence of $D_\parallel$ is detailed in Fig.~\ref{fig:dvsb}b and Tab.~\ref{tab:decay}. In all cases, an initial $\beta^0$-behavior
is followed by an algebraic decay, $D_\parallel\sim \beta^{- \alpha}$. The exponent $\alpha$ is smaller than unity for comparatively small $\Gamma$.
%and we expect convergence to zero for $\Gamma \ll 1$. 
With increasing $\Gamma$, however, $\alpha$ gradually increases and reaches $\alpha\approx 1$ at $\Gamma\approx\Gamma_\textrm{crit}(\beta)$. 
Thus, there exists Bohmian diffusion \emph{parallel to the field} which is another novel feature of SCPs.
Note that this rules out the predicted saturation $D_\parallel(\beta)/D_0\rightarrow 2/3$, for large  $\beta$ of Ref.~\cite{Cohen1984}.
%While their kinetic theory results (in particular those based on the modified Rostoker kernel) show reasonable agreement with the simulation results at small $\beta$ (open circles in Fig.~\ref{fig:dvsb}b), apparantly the high-field limit is incorrect. 
%

In conclusion, we have presented first-principle simulation results which, for the first time, illuminate the diffusion characteristics 
of SCPs under external magnetic fields in great detail, substantially generalizing results known for weakly coupled OCPs. 
At low coupling, three regimes of 
the magnetic-field dependence of $D_{\perp}$ have been identified, with a faster decay at intermediate B-fields. At large coupling parameters and strong magnetic fields we have 
found evidence of Bohmian $1/B$-diffusion indicating the dominant role of collective modes.
On the other hand, field-parallel diffusion in a SCP differs drastically from a weakly coupled plasma: $D_\parallel$ also depends on $B$, resulting in an algebraic decay,
 $D_\parallel \sim B^{-\alpha}$. This is a unique feature of high-$\Gamma$ liquid-like OCPs that is induced by many-particle correlations -- in particular shear and caging effects. Our results are expected to be of direct relevance for the SCP in compact stars and in inertial confinement fusion as well as for 
trapped ions and dusty plasmas. Moreover, they suggest to look for similar behavior of other transport quantities of magnetized SCP.

This work is supported by the Deutsche Forschungsgemeinschaft via SFB-TR 24 project A5 and a grant for computing time at the North-German Supercomputing Alliance (HLRN).

\end{document}